\documentclass[review]{elsarticle}
\usepackage{bm}
\usepackage{amssymb}
\usepackage{times}
\usepackage{latexsym}
\usepackage{graphicx}
\usepackage{color}
\usepackage{subfigure}
\usepackage{float}
\usepackage{amsmath}
\biboptions{numbers,sort&compress}

\begin{document}

\title{Minimal cooling speed for glass transition in a simple solvable energy landscape model}

\author[mymainaddress]{J. Quetzalc\'oatl Toledo-Mar\'in}
%%\ead[url]{www.elsevier.com}

\author[mysecondaryaddress1]{Isaac P\'erez Castillo}

\author[mymainaddress,mysecondaryaddress]{Gerardo G. Naumis\corref{mycorrespondingauthor}}
\cortext[mycorrespondingauthor]{Corresponding author}
\ead{naumis@fisica.unam.mx}

\address[mymainaddress]{Departamento de F\'{i}sica-Qu\'{i}mica, Instituto de
F\'{i}sica, Universidad Nacional Aut\'{o}noma de M\'{e}xico (UNAM),
Apartado Postal 20-364, 01000 M\'{e}xico, Distrito Federal,
M\'{e}xico}

\address[mysecondaryaddress1]{Departamento de Sistemas Complejos, Instituto de
F\'{i}sica, Universidad Nacional Aut\'{o}noma de M\'{e}xico (UNAM),
Apartado Postal 20-364, 01000 M\'{e}xico, Distrito Federal,
M\'{e}xico}

\address[mysecondaryaddress]{School of Physics Astronomy and Computational Sciences, George Mason University, Fairfax, Virginia 22030, USA }

\bibliographystyle{elsarticle-num}

\date{}

%\tableofcontents

\begin{abstract} 
The minimal cooling speed required to form a glass is obtained for a simple solvable energy landscape model. The model, made from a two-level system modified to include the topology of the energy landscape,  is able to capture either a glass transition or a crystallization depending on the cooling rate. In this setup, the minimal cooling speed to achieve  glass formation is then found to be related with the relaxation time and with the thermal history. In particular, we obtain that the thermal history encodes small fluctuations around the equilibrium population which are exponentially amplified near the glass transition, which mathematically corresponds to the boundary layer of the master equation. Finally, to verify our  analytical results, a kinetic Monte Carlo simulation was implemented. 
\end{abstract}

\maketitle

\section{Introduction}
The importance of glassy materials in our societies is indisputable. It is an essential component of numerous products that we use on daily basis, most often without noticing it. Even though the glass formation process has been extensively studied using different approaches, it remains an open and puzzling problem, and this far our best understanding of the process is barely limited at the phenomenological level \cite{naumis1998stochastic, phillips1996stretched, kerner2000stochastic, micoulaut1999glass, ramirez2011density, debenedetti1996metastable, debenedetti2001supercooled, stillinger1988supercooled, stillinger2002energy, Mauro1, trachenko2006stress, trachenko2008relationship}. The reason behind this situation is that glass formation is mainly a non-equilibrium process \cite{naumis2006variation}.

 From a fundamental and technological point of view, the most important variable for glass formation is the cooling speed \cite{Mauro1,Mauro2}. Indeed, the industrial use of metallic glasses
has been hampered for a while due to the high cooling speed required in order to form glasses \cite{metallicglass1,Angel1,Angel2}. However, by chemical modification, the cooling process of metallic glasses has been improved a lot \cite{metallicglass2}, and very recently it was possible to form a monocomponent metallic glass, achieved by hyperquenching \cite{metallicglass3}. Regarding the relationship between chemical composition and minimal cooling speed, Phillips  \cite{phillips1979topology} observed that for several chalcogenides, this minimal speed is a function of the rigidity. His initial observation was the starting point for an extensive investigation on the rigidity of glasses, yet this observation has not been quantitatively obtained in glass models although it is related with the energy landscape topology when the rigidity is taken into account \cite{naumisbifurcation, huerta, flores, naumistails}.

As the cooling rate effects on glass formation are poorly understood, one would expect that in any sensible model of glass transition, the phase transition to the crystal should be included for low cooling rates. However,  this point has been overlooked in several theories of glass formation. 
On the other hand, the energy landscape has been a useful picture to understand glass transition \cite{stillinger2002energy} but, due to its complicate high dimensional topology, it is difficult to understand how cooling rates are related with the topological sampling.

Simple models of glass transition have been introduced trying to  capture the physical properties of this phenomenon (see for instance \cite{Dyre1987,Dyre1995}). In particular, in a previous paper, a minimal simple solvable model of landscape that can display  either a crystalline phase or a glass transition depending on the cooling rate was presented  by one of us \cite{naumis}. Such model, a refinement of a  two-level system (TLS) model previously studied \cite{huse1986residual, langer1990nonequilibrium, langer1989entropy, langer1988entropy, brey1991residual}, included the most basic ingredients for a glass formation process: metastable states and the landscape topology \cite{naumis}.   As a result, the model was able to produce either a true phase transition or a glass transition in the thermodynamic limit \cite{naumis}. Nonetheless, there were important questions that were not tackled in our previous publication. In particular, it was not clear how to define a critical cooling speed that separates the transition either to a glass or to a crystal, and how this critical speed depends upon the physical characteristics of the system like relaxation times, energy barriers and the thermal history. In this study, we answer these open questions by obtaining analytical expressions to all these quantities. To verify these analytic calculations, a kinetic Monte Carlo is performed showing an excellent agreement. 

This article is organized as follows: section \ref{sec2} is devoted to recall the model and its features, as well as to obtain the system's behavior and an analytical expression of the glassy state when a given cooling protocol is applied \cite{naumis}. In section \ref{sec Char rel time}, we derive the characteristic relaxation time of our system. In section \ref{sec critical cooling rate} we obtain the relation between the metastable state, the cooling rate, the characteristic relaxation time and the thermal history of our system. In section \ref{sec KMC} we compare our results with  kinetic Monte Carlo simulation. Finally, in section \ref{sec conclusions} we summarize and discuss our findings.

\section{Revisiting a solvable energy landscape model: glass transition and crystallization}\label{sec2} 
\begin{figure}
\centering
\includegraphics[scale=0.6]{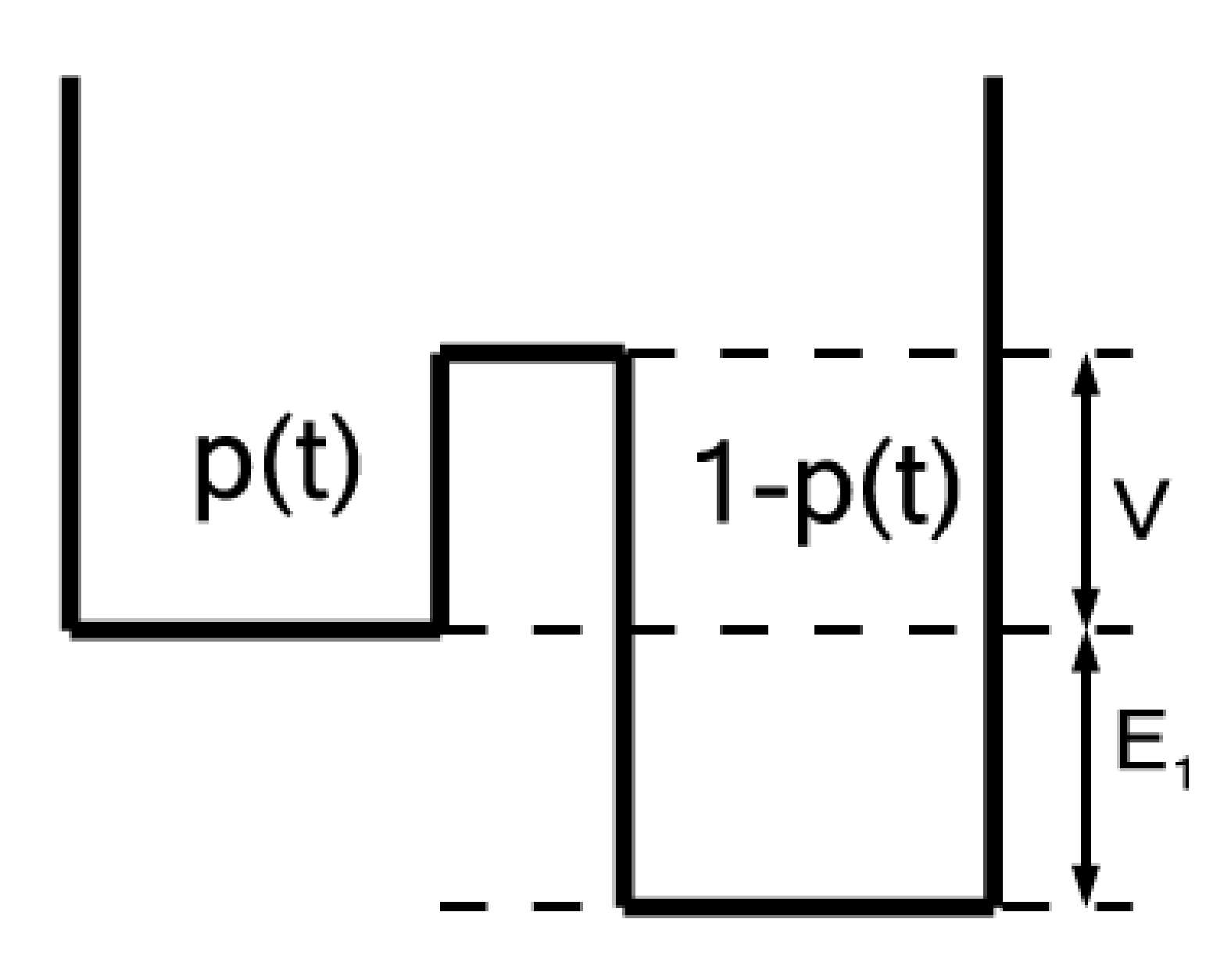}
\caption{The two level system energy landscape, showing the
barrier height $V$ and the asymmetry $E_1$ between the two levels. The
population of the upper well is $p(t)$ \cite{naumis}.}\label{system}
\end{figure}

The model is defined as follows: we have a two-level system (see figure \ref{system}) where state $0$ has energy $\epsilon_{0}=0$, and state 1 has energy $E_1=N \epsilon_{1}$ with degeneracy $g_{1}=2^{N}$. Hereafter $N$ corresponds to the number of particles in the system, and $g_{1}$ is just the complexity of the energy-landscape. 

When the system is in equilibrium at a certain temperature $T$,  the canonical partition function\footnote{From now on Boltzmann's constant $k_{B}=1$.} reads:
\begin{equation}
Z\left(T,N\right)=1+g_{1}e^{-E_{1}/T}\,,
\end{equation}
and the equilibrium probability  $p_0(T)$   to find the system in state $1$ is given by the usual ensemble average:
\begin{equation}
p_0(T)=\frac{g_{1}e^{-E_{1}/T}}{1+g_{1}e^{-E_{1}/T}}\label{solution in eq}\,.
\end{equation}
As shown in \cite{naumis}, for this equilibrium population the system experiences a phase transition associated with crystallization when the temperature crosses the critical value $T_c=\epsilon_1/\log(2)$.

To study the system out of equilibrium, one can assume \cite{naumis} a simple landscape topology in which all transition rates between metastable states are the same, and the transition rate from the metastable states to the ground state is also the same for all metasable states. In this setup, the probability $p(t)$   of finding  the system in one of the states with energy $E_1$ at time $t$ obeys the following master equation:
\begin{equation}\label{master equation}
\dot{p}(t)=-\Gamma_{10}p\left(t\right)+\Gamma_{01}g_{1}\left(1-p\left(t\right)\right) \,,
\end{equation}
where $\Gamma_{10}$ (resp. $\Gamma_{01}$) corresponds to the transition probability per time of going from state $1$ to state $0$ (resp. state $0$ to state $1$). Detailed balance condition yields:
\begin{equation}\label{15}
\frac{\Gamma_{01}}{\Gamma_{10}}=e^{-E_{1}/T}
\end{equation}
and $\Gamma_{10}=\Gamma_{0}e^{-V/T}$, where $V$ is the height of the barrier wall between state $1$ and state $0$, and $\Gamma_{0}$ is a small frequency of oscillation at the bottom of the walls.

Now we are interested in the process of arresting the system in one of the higher energy states by a rapid cooling, as it  happens with glasses.  In particular, we are interested in studying the system as the  temperature goes from $T>T_c$ to $T=0$ by a cooling rate determined by a given protocol $T(t)$. Notice that since $T=T(t)$, the population described by Eq. (\ref{master equation}) will be denoted at times by $p(T)$, not to be confused with the equilibrium probability $p_0(T)$.  Experimentally, a linear cooling is usually used. However, for the purposes of the model, it is much simpler to use a hyperbolic cooling protocol $T(t)=T_0/(1+Rt)$, where $T_0$ is the initial temperature at which the system is in equilibrium and $R$ is the cooling rate. The results using both protocols are similar since basically the equations can be approximated using the boundary layer theory of differential equations \cite{langer1988entropy,naumis}. By boundary layer, we mean that in Eq. (\ref{master equation}), the time derivative can be neglected above $T_c$ and the system behaves as an equilibrated system. However, as $T \rightarrow T_c$, the derivative can not be longer neglected, since its order is similar to the other terms. A similar situation happens with the Navier-Stokes equations in fluids, which are reduced to Euler equations far from the boundary, but near the boundary the full equation is needed, producing effects like turbulence. 

The solution to the master equation (\ref{master equation}) given the cooling protocol is:
\begin{equation} \label{Gen solution}
 \begin{aligned}
P\left(x,\delta\right)=e^{\frac{1}{\delta}\left(x+g_{1}\frac{x^{\mu+1}}{\mu+1}\right)}\left(P(0,\delta) -\frac{g_1}{\delta}\int_0^xdyy^\mu e^{-\frac{1}{\delta}\left(y+g_1\frac{y^{\mu+1}}{\mu+1}\right)}\right)\,,
\end{aligned}
\end{equation}
where $p(t)=P\left(x(t),\delta\right)$ with $x(t)=\exp(-V/T(t))$,  $\delta=RV/\Gamma_0T_0$ is the dimensionless cooling rate, and the parameter $\mu=E_1/V$ measures the asymmetry of the well. 

\begin{figure}
\centering
\includegraphics[scale=0.4]{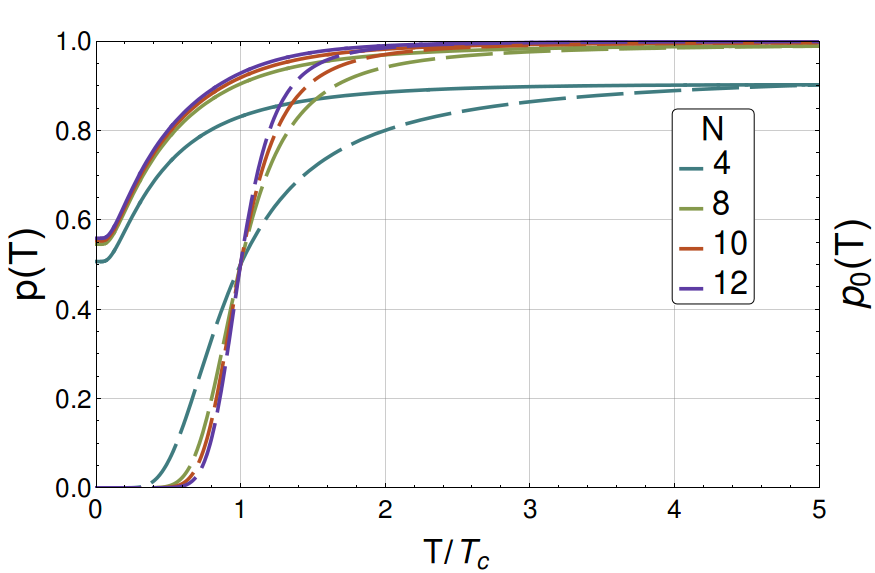}
\caption{$p(T)$ in equilibrium (dashed lines) and non-equilibrium (solid lines) with a hyperbolic colling protocol. The parameters were fixed at $T_0=5T_c, V=0.5, R=20$ and $\Gamma_0=1$}\label{peqvsp}
\end{figure}

\begin{figure}
\centering
\includegraphics[scale=0.4]{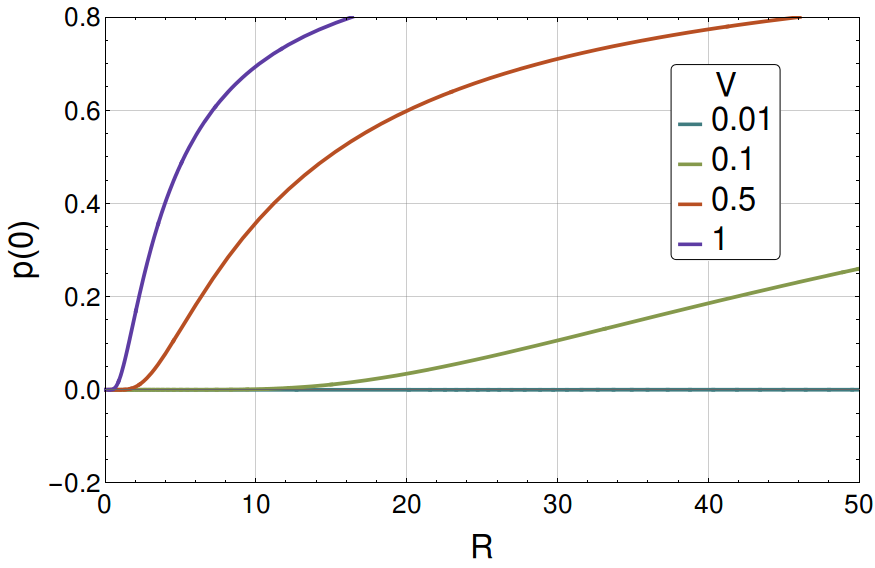}
\caption{$p(T=0)$ for different barrier heights. The parameters were fixed at $T_0=5T_c,\, N=500$ and $\Gamma_0=1$}\label{p0tv}
\end{figure}

According to equations (\ref{solution in eq}) and (\ref{Gen solution}), and as  we  can appreciate in figure \ref{peqvsp}  for different number of particles $N$, when the systems is cooled down to $T=0$ there is a residual population, i.e., $p(T=0)\neq0$ ($P(x=0,\delta)\neq 0$) indicative of a glassy behavior due to the trapping of the system in a metastable state \cite{langer1989entropy,naumis}. In fact, we can obtain an analytical expression for $P(x=0,\delta)$ from Eq. (\ref{Gen solution}) by assuming that the system is initially in thermal equilibrium at temperature $T_0>T_c$ before being cooled, viz.
\begin{equation} \label{p0}
 \begin{aligned}
P\left(0,\delta\right)=\frac{g_1x_0^\mu}{1+g_1x_0^\mu}e^{-\frac{1}{\delta}\left(x_0+g_{1}\frac{x_0^{\mu+1}}{\mu+1}\right)}+\frac{g_1}{\delta}\int_0^{x_{0}}dyy^\mu e^{-\frac{1}{\delta}\left(y+g_1\frac{y^{\mu+1}}{\mu+1}\right)},
\end{aligned}
\end{equation}
with $x_0=\exp(-V/T_0)$. As we can see in figure \ref{p0tv} , the residual population given by Eq.  (\ref{p0}) has a strong dependence of the barrier height and the cooling rate.

\section{Characteristic relaxation times of the model}\label{sec Char rel time}
Let us focus now on quantifying the dependence of this residual population on the energy landscape.  In particular we would like to have a criterion to discern \textit{how fast one should cool the system down to obtain a residual population}. \\
Clearly, in order to trap the system the cooling must be such that the system does not have enough time to reach equilibrium, so let us first determine the characteristic relaxation times of the system. To do so, we take  the parameters of the model to be fixed but the system is not in equilibrium, i.e., the temperature is fixed and the system is perturbed in such a way that at $t=0$, the population is $p(t=0)=\rho$, where $\rho$ takes values between $0$ and $1$.  Looking from the master equation  (\ref{master equation})  and the detailed balance condition (\ref{15}) how the system relaxes towards $p_0(T)$,  we obtain an exponential decay:
\begin{equation}
p(t)=p_0(T)+\left(\rho-p_0(T)\right)\exp(-t/\tau))\label{solution returning to eq}\,,
\end{equation}
from which we define the characteristic relaxation time $\tau=1/(\Gamma_{10}+\Gamma_{01}g_1)$ .

\begin{figure}
\centering
\includegraphics[scale=0.4]{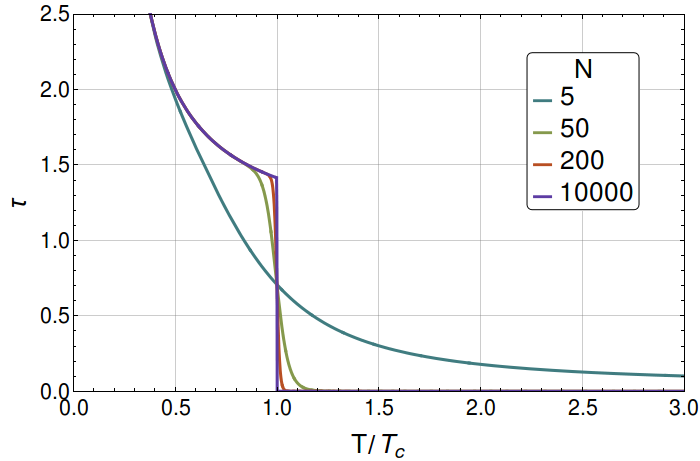}
\caption{Characteristic relaxation time $\tau$ as function of the number of particles $N$, with fixed parameters $V=0.5,\epsilon_1=1, \Gamma_0=1$. }\label{chartime}
\end{figure}

Notice that for $N\gg1$  the characteristic relaxation time (\ref{solution returning to eq}) goes as $\sim(g_1\Gamma_{01})^{-1}$ for $T>T_c$,  while for $T<T_c$ goes as $\sim(\Gamma_{10})^{-1}$ (see figure  \ref{chartime}). In particular, when $T$ crosses the critical temperature $T_c$, the $\tau$ has a jump of height $\tau\simeq\exp(-V\log(2)/\epsilon_1)$. Hence, when $T>T_c$ and the system is in state $0$ the transition time is virtually zero, whereas when $T<T_c$ and the system is in state $1$ the transition time grows exponentially with $V/T$.

\section{Critical cooling rate and glass transition} \label{sec critical cooling rate}
\begin{figure}
\centering
\includegraphics[scale=0.4]{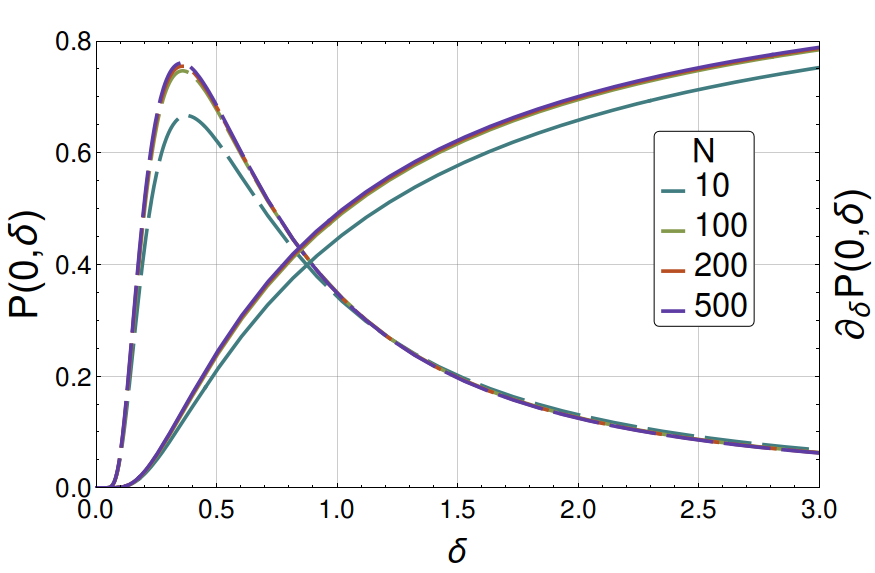}
\caption{Here we plot $P(0,\delta)$ (continous lines) and $\frac{\partial{P(0,\delta)}}{\partial\delta}$ (dashed lines) as function of $\delta$ for different number of particles $N$. We fixed the parameters at $T_0=5T_c, \Gamma_0=1, V=0.5, \epsilon_1=1$}\label{p0x&dp0x}
\end{figure}

As we have seen above for a certain cooling rate $R$ there is a non-zero probability of finding the system in state $1$ at $T=0$, and the time needed for the system to transition from state $1$ to state $0$ goes as $\exp(V/T)$ when $T<T_c$. We would like now to have  simple criterion that relates the cooling rate and a substantial residual population indicative of a glassy behavior. Noticing that Eq. (\ref{p0}) is continuous and reaches zero only when $\delta =0$, we then take as a criterion the inflection point of $p(0,\delta)$ (as shown in figure \ref{p0x&dp0x}). Thus by denoting $\delta_c$ the cooling rate at the inflection point, we can associate a strong glass forming tendency for $\delta>\delta_c$.    

\begin{figure}
\centering
\includegraphics[scale=0.4]{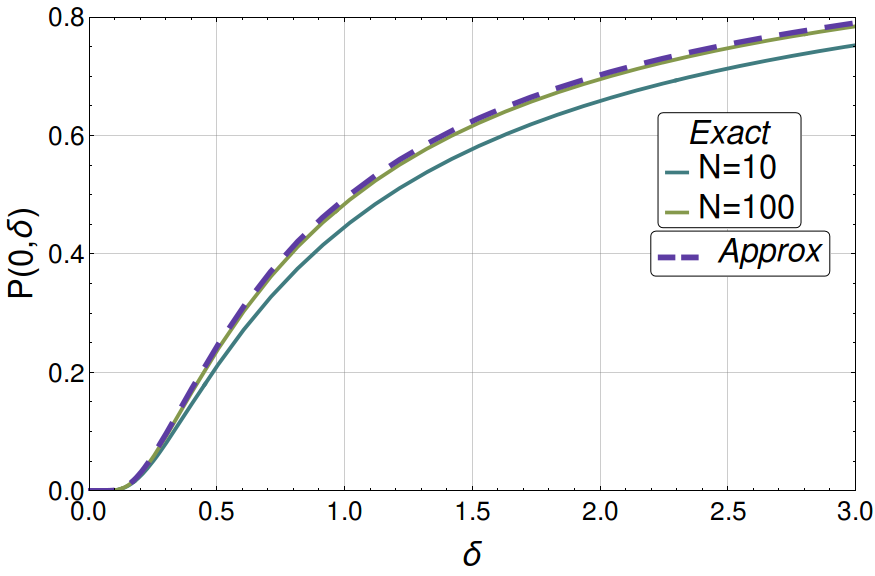}
\caption{Comparison between Eq. (\ref{I2 approx}) and Eq. (\ref{p0}) as function of $\delta$. The parameters were fixed at $V=0.5, \epsilon_1=1, T_0=5T_c, \Gamma_0=1, $}\label{comparison p0 and p0 approx}
\end{figure}

To find the dependence of $\delta_c$ as a function of the parameters of the model we proceed as follows. We write Eq. (\ref{p0}) as $P(0;\delta,\mu,N)=I_1+I_2$, where:
\begin{equation}\label{I1}
\begin{aligned}
I_1=\frac{g_1x_0^\mu}{1+g_1x_0^\mu}\exp\left[-\frac{1}{\delta}\left(x_0+g_{1}\frac{x_0^{\mu+1}}{\mu+1}\right)\right]\,,
\\
I_2=\frac{g_1}{\delta}\int_0^{x_{0}}dyy^\mu\exp\left[-\frac{1}{\delta}\left(y+g_1\frac{y^{\mu+1}}{\mu+1}\right)\right]\,.
\end{aligned}
\end{equation}
Integrating the expression of $I_2$ in equation (\ref{I1}) by parts leads to
\begin{equation}\label{I2}
\begin{aligned}
I_2=1-e^{-\frac{1}{\delta}\left(x_0+g_1\frac{x_0^{\mu+1}}{\mu+1}\right)}-\int_0^{x_0}\frac{dy}{\delta} e^{-\frac{y}{\delta}\left(1+g_1\frac{y^\mu}{\mu+1}\right)}\,.
\end{aligned}
\end{equation}
Let us denote $x_c=x(T_c)$. In the thermodynamic limit where $N\gg1$, for $y<x_c$ we have that $g_1\frac{y^\mu}{\mu+1}\simeq0$, whereas if $y>x_c$ results in $g_1\frac{y^\mu}{\mu+1}\gg1$. Thus, we may approximate the last term in expression (\ref{I2}) as:
\begin{equation}\label{I2 approx}
-\int_0^{x_0}\frac{dy}{\delta}e^{-\frac{y}{\delta}\left(1+g_1\frac{y^\mu}{\mu+1}\right)}\simeq e^{-y/\delta}\left.\vphantom\int{}\right|_0^{x_c}\,.
\end{equation}
Thus, substituting equations (\ref{I2}) and (\ref{I2 approx}) in Eq. (\ref{p0}) yields:
\begin{equation} \label{p0 approx}
 \begin{aligned}
P\left(0,\delta\right)\simeq&-\frac{1}{1+g_1x_0^\mu}e^{-\frac{1}{\delta}\left(x_0+g_{1}\frac{x_0^{\mu+1}}{\mu+1}\right)}+\exp\left(-\frac{x_c}{\delta}\right)\,.
\end{aligned}
\end{equation}
Since $x_0>x_c$, then Eq. (\ref{p0 approx}) can be approximated as:
\begin{equation}
P\left(0;\delta\right)\simeq\exp\left(-\frac{x_c}{\delta}\right) \label{p0 approxx}\,.
\end{equation}
Finally, writing Eq. (\ref{p0 approxx}) in terms of $R$ yields:
\begin{equation}
p(T=0)\approx \exp\left(-\frac{T_0\Gamma_{10}^{c}}{RV}\right)	 \label{p0 approxx2}\,,
\end{equation}
where 
\begin{equation}
\Gamma_{10}^{c}=\Gamma_0 e^{-V/T_c}\,.
\end{equation}
In figure \ref{comparison p0 and p0 approx} we have compared the exact result and the approximation of $P(0,\delta)$ (Eqs. \ref{p0} and \ref{p0 approxx}). We can clearly appreciate how the exact results tends to our approximation Eq. (\ref{p0 approxx2}) as $N$ increases. Notice that expression (\ref{p0 approxx2}) relates the residual population with the cooling rate $R$ and the characteristic time $\tau$ in a very simple and intuitive manner.  This result tells us that trapping the system in the metastable state ultimately depends on the cooling rate solely applied in a region close to the phase transition zone \cite{prados1997dynamical}, although there is a catch.

 Suppose that we cool the system starting from $T_1$ with a cooling rate $R_1$, and we repeat the process starting from $T_2\neq T_1$ with a cooling rate $R_2\neq R_1$. The residual population $p(0)$ may be the same in both cases provided $T_1/R_1=T_2/R_2$. This implies that if $T_1>T_2$ then $R_1>R_2$, i.e., to trap the system in state $1$ starting from $T_1$ we would need a cooling rate $R_1$ bigger than the one needed if the cooling started at $T_2<T_1$. Thus, we would be compelled to assume that the "best" way to trap the system in our model would be to set the initial temperature $T_0$ as close as possible to $T_c$. However, in our model $T_0$ is the initial temperature in which the system is in thermodynamical equilibrium. Figure \ref{peqvsp} illustrates this idea, i.e., even though the transition occurs in $T_c$ the non-equilibrium system's path differs from the equilibrium system's path even before reaching $T_c$, therefore there is a lower bound for $T_0$. This means that the thermal history encodes small fluctuations around the equilibrium population which are exponentially amplified near the glass transition. This region of the glass transition corresponds precisely to the boundary layer limit.

Finally, using our approximation Eq. (\ref{p0 approxx}) we can define the critical dimensionless cooling rate $\delta_c=x_c/2$ that gives the inflection point of $P(0,\delta)$ as a function of $\delta$. Evaluating $\delta_c$ in our approximation (Eq. \ref{p0 approxx}) gives always the same population at the inflection point $P(0,\delta_c)=e^{-2}\approx 0.13$. This means that below $\delta_c$ there is a probability of a residual population lower than $\sim 0.13$.  

\section{Kinetic Monte Carlo simulation} \label{sec KMC}
\begin{figure}
\centering
\includegraphics[scale=0.4]{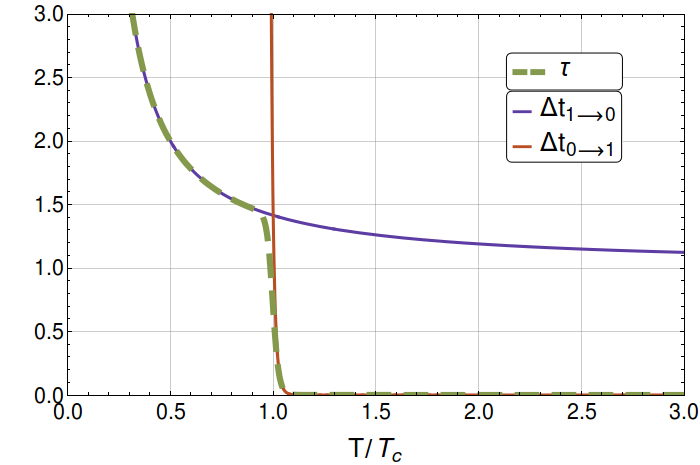}
\caption{KMC transition times (continous lines) and characteristic relaxation time $\tau$ obtain from Eq. \ref{solution returning to eq} (dashed line). We have fixed the parameters to $T_0=5T_c, V=0.5, \epsilon_1=1, R=0.01, N=100$}\label{KMCtranstimes}
\end{figure}

To asses the validity of our mathematical analysis  we have compared our analytic results with a Kinetic Monte Carlo (KMC) simulation. The simulation was done in a standard way (see for instance \cite{prados1997dynamical}). The (residence) time $\Delta t_{ij}$ the system spends in state $i$ $(i,j=\lbrace0,1\rbrace,\,i\neq j)$, given the frozen state condition is not fulfilled (\cite{prados1997dynamical}), is determined by the relation:
\begin{equation}\label{relation}
-\log(x)=\int_t^{t+\Delta t_{ij}}dt'W_{ij}(t')\,,
\end{equation}
with
\begin{eqnarray}
\begin{array}{c}\label{Wij}
W_{10}(t)=\Gamma_{10}(t)\,, \\ 
W_{01}(t)=g_1\Gamma_{01}(t)\,,
\end{array} 
\end{eqnarray}
and $x$  a uniformly distributed random number between $0$ and $1$. Thus, from relation (\ref{relation}) and expressions (\ref{Wij}) we obtain:
\begin{equation}
\Delta t_{1 0}=-\frac{T_0}{VR}\log\left(1+\frac{\log(x)VR}{\Gamma_0 T_0}e^{V(1+Rt)/T_0}\right) \label{dt10}\,,
\end{equation}
\begin{equation}
\begin{aligned}
\Delta t_{0 1}=-\frac{T_0}{(V+E_1)R}\log\left(1+\frac{\log(x)(V+E_1)R}{\Gamma_0 T_0}\times\right.\\\left.\exp\left(-N\log(2)+(V+E_1)(1+Rt)/T_0\right) \vphantom{\sum}\right) \label{dt01}\,.
\end{aligned}
\end{equation}
In figure \ref{KMCtranstimes} we have plotted expressions (\ref{dt01}) and (\ref{dt10}) for $x=e^{-1}$ and a small cooling rate $R$, i.e., a quasi-equilibrium cooling rate, and we  have compared it with the characteristic relaxation time $\tau$ as function of $T$. Notice that when $T>T_c$, $\Delta t_{10}>\Delta t_{01}$, whereas when $T<T_c$ results in $\Delta t_{10}<\Delta t_{01}$. Furthermore, when $T>T_c$ the residence time $\Delta t_{10}$ corresponds to the system's characteristic relaxation time, while when $T<T_c$ the residence time $\Delta t_{01}$ correspond to the system's characteristic relaxation time.

\begin{figure}
\centering
\includegraphics[scale=0.4]{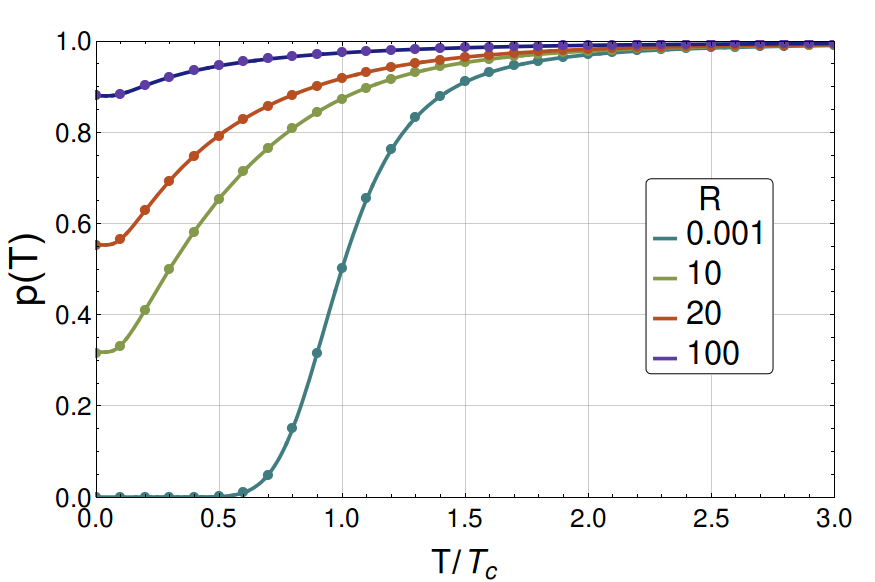}
\caption{$p(T)$. Comparison between our exact solution (continous line) and our KMC simulation (points), for different cooling rates $R$ and the following choice of parameters: $N=10,\,T_0=5T_c,\,V=0.5,\,\Gamma_0=1$. The KMC simulation was done with an ensemble of $10^5$ systems. }\label{KMC&AnSol10part}
\end{figure}

\begin{figure}
\centering
\includegraphics[scale=0.4]{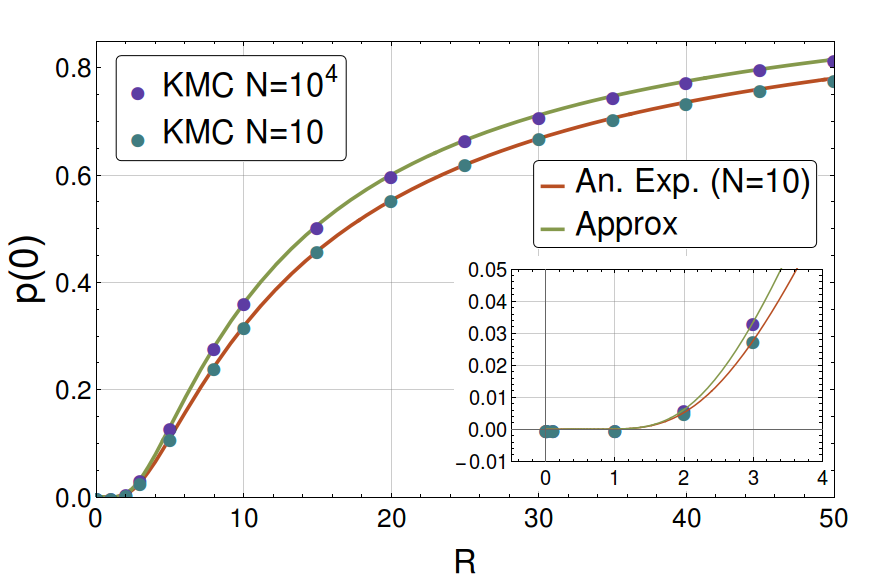}
\caption{$p(0)$. Comparison between our exact and approximate expression (continous line) with our KMC simulation (points) for different cooling rates $R$ and the following choice of parameters: $N=10,\,T_0=5T_c,\,V=0.5,\,\Gamma_0=1$. The KMC simulation was done with an ensemble of $10^5$ systems and $N=\lbrace10,10^4\rbrace$.}\label{KMCp0}
\end{figure}

In figure \ref{KMC&AnSol10part} we  have compared $p(T)$ (Eq. \ref{Gen solution} as function of $T(t)$) with our KMC simulation for different cooling rates. As for figure \ref{KMCp0}, we have compared Eq. (\ref{p0 approxx}) with our KMC simulation. The match between our analytical results and the KMC simulation is outstanding. We should stress the fact that the computational cost by the KMC simulation is much less than the numerical evaluation of $p(T)$ for large $N$.

Following  \cite{prados1997dynamical}, given that the system is initially in state $i$, the probability that it will remain frozen in this state forever is $\exp(-s_i^{(\infty)})$  with :
\begin{equation}
s_i^{(\infty)}\equiv \lim_{t\rightarrow \infty}\int_0^t dt' W_{ij}(t')\,.
\end{equation}
In our model this implies that
\begin{eqnarray}
\exp\left(-s_1^{(\infty)}\right)=\exp\left(-\frac{x_0}{\delta}\right) \label{KMC relation1}\,, \\
\exp\left(-s_0^{(\infty)}\right)=\exp\left(-\frac{g_1 x_0^{\mu+1}}{\delta(\mu+1)}\right) \label{KMC relation2}\,.
\end{eqnarray}

Notice that at $T_0=T_c$, the expression (\ref{KMC relation1}) is the same as our approximation of $P(0;\delta)$ given by Eq. \ref{p0 approxx}. Therefore, trapping the system in state $1$ ultimately depends on doing so at the transition point, although the system's path towards that transition point is relevant. Hence, the system has thermal history.

\section{Conclusions} \label{sec conclusions}
Using a simple energy landscape model that shows a phase transition and a glass transition depending on the cooling rate, we found a relation between the residual population, the characteristic relaxation times, the cooling rate and the thermal history. In particular, the residual population, which is a measure of the glass forming tendency, turns out to have an inflection point as a function of the cooling rate. This allows to define a critical cooling rate in the sense that higher cooling speeds than the critical one result in an increased glass forming tendency. The critical rate depends upon the relaxation time for crystallization, the phase transition temperature and the thermal history. Interestingly, the thermal history produces small fluctuations around the 
equilibrium population which are exponentially amplified near the glass transition,  which in fact corresponds to the region of the master equation boundary layer. In other words, the thermal history encodes the sensibility to the initial conditions of the system, as happens  with turbulence inside the boundary layer.   
Finally, a kinetic Monte Carlo simulation was performed to check the analytically obtained residual populations 
and the relaxation times. An excellent agreement was found between both methods. In fact, the relaxation time is a nice interpolation of the residence times obtained from the Monte Carlo. All these results could be used for more realistic energy landscapes, by using connectivity maps \cite{Wales1,Wales2,Wales3}.

\bibliography{mybib}

\begin{thebibliography}{10}
\expandafter\ifx\csname url\endcsname\relax
  \def\url#1{\texttt{#1}}\fi
\expandafter\ifx\csname urlprefix\endcsname\relax\def\urlprefix{URL }\fi
\expandafter\ifx\csname href\endcsname\relax
  \def\href#1#2{#2} \def\path#1{#1}\fi

\bibitem{naumis1998stochastic}
G.~G. Naumis, R.~Kerner, Stochastic matrix description of glass transition in
  ternary chalcogenide systems, Journal of non-crystalline solids 231~(1)
  (1998) 111--119.

\bibitem{phillips1996stretched}
J.~Phillips, Stretched exponential relaxation in molecular and electronic
  glasses, Reports on Progress in Physics 59~(9) (1996) 1133.

\bibitem{kerner2000stochastic}
R.~Kerner, G.~G. Naumis, Stochastic matrix description of the glass transition,
  Journal of Physics: Condensed Matter 12~(8) (2000) 1641.

\bibitem{micoulaut1999glass}
M.~Micoulaut, G.~Naumis, Glass transition temperature variation, cross-linking
  and structure in network glasses: a stochastic approach, EPL (Europhysics
  Letters) 47~(5) (1999) 568.

\bibitem{ramirez2011density}
P.~E. Ram{\'\i}rez-Gonz{\'a}lez, L.~L{\'o}pez-Flores, H.~Acu{\~n}a-Campa,
  M.~Medina-Noyola, Density-temperature-softness scaling of the dynamics of
  glass-forming soft-sphere liquids, Physical review letters 107~(15) (2011)
  155701.

\bibitem{debenedetti1996metastable}
P.~G. Debenedetti, Metastable liquids: concepts and principles, Princeton
  University Press, 1996.

\bibitem{debenedetti2001supercooled}
P.~G. Debenedetti, F.~H. Stillinger, Supercooled liquids and the glass
  transition, Nature 410~(6825) (2001) 259--267.

\bibitem{stillinger1988supercooled}
F.~H. Stillinger, Supercooled liquids, glass transitions, and the kauzmann
  paradox, The Journal of chemical physics 88~(12) (1988) 7818--7825.

\bibitem{stillinger2002energy}
F.~H. Stillinger, P.~G. Debenedetti, Energy landscape diversity and supercooled
  liquid properties, The Journal of chemical physics 116~(8) (2002) 3353--3361.

\bibitem{Mauro1}
M.~M. Smedskjaer, J.~C. Mauro, Y.~Yue, Prediction of glass hardness using
  temperature-dependent constraint theory, Physical review letters 105~(11)
  (2010) 115503.

\bibitem{trachenko2006stress}
K.~Trachenko, A stress relaxation approach to glass transition, Journal of
  Physics: Condensed Matter 18~(19) (2006) L251.

\bibitem{trachenko2008relationship}
K.~Trachenko, C.~Roland, R.~Casalini, Relationship between the
  nonexponentiality of relaxation and relaxation time in the problem of glass
  transition, The Journal of Physical Chemistry B 112~(16) (2008) 5111--5115.

\bibitem{naumis2006variation}
G.~G. Naumis, Variation of the glass transition temperature with rigidity and
  chemical composition, Physical Review B 73~(17) (2006) 172202.

\bibitem{Mauro2}
J.~C. Mauro, D.~C. Allan, M.~Potuzak, Nonequilibrium viscosity of glass,
  Physical Review B 80~(9) (2009) 094204.

\bibitem{metallicglass1}
A.~Inoue, Stabilization of metallic supercooled liquid and bulk amorphous
  alloys, Acta materialia 48~(1) (2000) 279--306.

\bibitem{Angel1}
J.~Reyes-Retana, G.~Naumis, The effects of si substitution on the glass forming
  ability of ni--pd--p system, a dft study on crystalline related clusters,
  Journal of Non-Crystalline Solids 387 (2014) 117--123.

\bibitem{Angel2}
J.~Reyes-Retana, G.~Naumis, Ab initio study of si doping effects in pd--ni--p
  bulk metallic glass, Journal of Non-Crystalline Solids 409 (2015) 49--53.

\bibitem{metallicglass2}
M.~Ashby, A.~Greer, Metallic glasses as structural materials, Scripta
  Materialia 54~(3) (2006) 321--326.

\bibitem{metallicglass3}
L.~Zhong, J.~Wang, H.~Sheng, Z.~Zhang, S.~X. Mao, Formation of monatomic
  metallic glasses through ultrafast liquid quenching, Nature 512~(7513) (2014)
  177--180.

\bibitem{phillips1979topology}
J.~C. Phillips, Topology of covalent non-crystalline solids i: Short-range
  order in chalcogenide alloys, Journal of Non-Crystalline Solids 34~(2) (1979)
  153--181.

\bibitem{naumisbifurcation}
G.~Naumis, J.~Phillips, Bifurcation of stretched exponential relaxation in
  microscopically homogeneous glasses, Journal of Non-Crystalline Solids
  358~(5) (2012) 893--897.

\bibitem{huerta}
A.~Huerta, G.~Naumis, Relationship between glass transition and rigidity in a
  binary associative fluid, Physics Letters A 299~(5) (2002) 660--665.

\bibitem{flores}
H.~M. Flores-Ruiz, G.~G. Naumis, J.~Phillips, Heating through the glass
  transition: A rigidity approach to the boson peak, Physical Review B 82~(21)
  (2010) 214201.

\bibitem{naumistails}
G.~Naumis, G.~Cocho, The tails of rank-size distributions due to multiplicative
  processes: from power laws to stretched exponentials and beta-like functions,
  New Journal of Physics 9~(8) (2007) 286.

\bibitem{Dyre1987}
J.~C. Dyre, Master-equation approach to the glass transition, Physical Review
  Letters 58~(8) (1987) 792.

\bibitem{Dyre1995}
J.~C. Dyre, Energy master equation: a low-temperature approximation to
  b\"assler's random-walk model, Physical Review B 51~(18) (1995) 12276.

\bibitem{naumis}
G.~G. Naumis, Simple solvable energy-landscape model that shows a thermodynamic
  phase transition and a glass transition, Physical Review E 85~(6) (2012)
  061505.

\bibitem{huse1986residual}
D.~A. Huse, D.~S. Fisher, Residual energies after slow cooling of disordered
  systems, Physical review letters 57~(17) (1986) 2203.

\bibitem{langer1990nonequilibrium}
S.~A. Langer, J.~P. Sethna, E.~R. Grannan, Nonequilibrium entropy and entropy
  distributions, Physical Review B 41~(4) (1990) 2261.

\bibitem{langer1989entropy}
S.~A. Langer, A.~T. Dorsey, J.~P. Sethna, Entropy distribution of a two-level
  system: An asymptotic analysis, Physical Review B 40~(1) (1989) 345.

\bibitem{langer1988entropy}
S.~A. Langer, J.~P. Sethna, Entropy of glasses, Physical review letters 61~(5)
  (1988) 570.

\bibitem{brey1991residual}
J.~J. Brey, A.~Prados, Residual properties of a two-level system, Physical
  Review B 43~(10) (1991) 8350.

\bibitem{prados1997dynamical}
A.~Prados, J.~Brey, B.~S{\'a}nchez-Rey, A dynamical monte carlo algorithm for
  master equations with time-dependent transition rates, Journal of statistical
  physics 89~(3-4) (1997) 709--734.

\bibitem{Wales1}
T.~F. Middleton, J.~Hern{\'a}ndez-Rojas, P.~N. Mortenson, D.~J. Wales, Crystals
  of binary lennard-jones solids, Physical Review B 64~(18) (2001) 184201.

\bibitem{Wales2}
T.~F. Middleton, D.~J. Wales, Energy landscapes of some model glass formers,
  Physical Review B 64~(2) (2001) 024205.

\bibitem{Wales3}
V.~K. de~Souza, D.~J. Wales, Connectivity in the potential energy landscape for
  binary lennard-jones systems, The Journal of chemical physics 130~(19) (2009)
  194508.

\end{thebibliography}

\end{document}